\documentclass[12pt]{article}
  \usepackage{amsthm,amsfonts}
  \usepackage{amsmath}
\newcommand{\bea}   {\begin{eqnarray}}
\newcommand{\eea}   {\end{eqnarray}}
\usepackage{subeqnarray}
\usepackage[brazilian]{babel}
\usepackage[latin1]{inputenc}
\begin{document}
\renewcommand{\thefootnote}{\fnsymbol{footnote}}
\date{}

\begin{center}
{\bf Sull'origine della condizione di quantizzazione delle cariche dei dioni.}
\end{center}

\begin{center}
{\it F. Caruso\footnote{In congedo dall'Instituto de F\'{\i}sica da Universidade do Estado do Rio de Janeiro, Brasile.}}\\
Istituto di Fisica dell'Universit\`{a} di Torino\\
Istituto Nazionale di Fisica Nucleare -- Sezione di Torino
\end{center}
\setcounter{footnote}{0}

\section{Introduzione}
\renewcommand{\thefootnote}{\arabic{footnote}}

Quando si discute della natura elettrica o magnetica di una carica, nel contesto della teoria elettromagnetica (classica) di Maxwell si ha bisogno
di {\it supporre} l'esistenza di densit\`{a} di cariche e correnti di origine magnetica\footnote{J.D. JACKSON, {\it Classical Electrodynamics}, second edizione (New York, N.Y., 1975).}, $\rho_m$ e $j_m$, oltre a
quelle di origine elettrica $\rho_e$ e $j_e$. In tal caso, le equazioni di Maxwell vengono generalizzate nel modo seguente:
\begin{equation}
\left\{
\begin{array}{l}
\nabla \cdot D=4\pi \rho_e\ ,\\
\\
\nabla \times H=\displaystyle{\frac{1}{c}}\ \displaystyle{\frac{\partial D}{\partial t}}+\displaystyle{\frac{4\pi}{c}}\ j_e;\\
\\
\nabla \cdot B = 4\pi \rho_m,\\\
\\
-\nabla \times E=\displaystyle{\frac{1}{c}}\ \displaystyle{\frac{\partial B}{\partial t}}+\displaystyle{\frac{4\pi }{c}}\ j_m,\\
\end{array}
\right.
\label{eq1}
\end{equation}
L'insieme delle equazioni ottenute con questa ipotesi \'{e} invariante per le trasformazioni ortogonali dette ``trasformazioni duali dei campi'',
\begin{equation}
\left\{
\begin{array}{l}
E=E^{\prime}\cos \theta +H^{\prime}\sin \theta,\\
\\
D=D^{\prime}\cos \theta +B^{\prime}\sin \theta,\\
\\
H=-E^{\prime}\sin\theta +H^{\prime}\cos \theta,\\
\\
B=-D^{\prime}\sin \theta +B^{\prime}\cos \theta.\\
\end{array}
\right.
\label{eq2}
\end{equation}
\textit{se} le densit\`{a} di cariche e correnti si trasformano nella stessa maniera:
\begin{equation}
\left\{
\begin{array}{l}
\rho_e=\rho^{\prime}_e\cos \theta +\rho^{\prime}_m\sin \theta,\\
\\
j_e=j^{\prime}_e\cos \theta +j^{\prime}_m\sin \theta;\\
\\
\rho_m=p^{\prime}_e\sin \theta +\rho^{\prime}_m\cos \theta,\\
\\
j_m=j^{\prime}_e\sin \theta +j^{\prime}_m\cos \theta.\\
\end{array}
\right.
\label{eq3}
\end{equation}
Questa necessit\`{a} (il se corsivo) discende naturalmente dal fatto che la teoria di Maxwell considera i campi e le particelle come entit\`{a} fondamentali. Ci sono, peraltro, due difficolt\`{a} che sono in relazione con questo fatto e con queste trasformazioni. La prima \`{e} la difficolt\`{a} di definirle per i campi nei punti dove essi sono singolari -- cio\`{e} dove sono le cariche -- poich\'{e} i campi divergono in questi punti. L'altra difficolt\`{a} \`{e} dovuta al fatto che le cariche e le correnti sono considerate come attributi delle particelle. Ci\`{o} genera la necessit\`{a} di trovare un argumento con cui si possa concludere che il rapporto fra la carica magnetica e quella elettrica di tutte le particelle sia universale poich\`{e}, solo cos\'{\i}, \`{e} possibile scegliere una trasformazione duale particolare a causa della quale le densit\`{a} di cariche e correnti magnetiche risultino nulle.

In questo caso le equazioni di Maxwell assumono la loro forma usuale, riassumendo tutta la fenomenologia classica conosciuta, senza la necessit\`{a}  d'introdurre cariche magnetiche.

Malgrado le difficolt\`{a} di cui sopra, la generalizzazione formale (1) delle equazioni di Maxwell permette di aprire una
discussione interessante su certe propriet\`{a} o caratteristiche delle particelle cariche. In primo luogo sorge il problema dell'universalit\`{a}, o meno,  del rapporto $G/E$ fra la carica magnetica ed elettrica di una stessa particella; non sembra peraltro possibile trovare un argomento entro la stessa teoria di Maxwell che possa dare una soluzione a tale problema. Resta la possibilit\`{a} di affrontare il problema nell'ambito della natura quantistica della carica elettrica che porta anche alla possibilit\`{a} dell'esistenza degli oggetti chiamati dioni ({\it dyons}) che possiedono entrambe le cariche elettrica e magnetica. L'idea di Dirac\footnote{P.A.M. DIRAC; {\it Proc. R. Soc. London Ser. A}, {\bf 133}, 60 (1931); {\it Phys. Rev}., {\bf 74}, 817 (1948).}  che la quantizzazione della carica elettrica \`{e} legata all'esistenza di un monopolo in natura \`{e} ormai abbastanza difusa. La carica elettrica e quella magnetica sarebbero legate dall'equazione
\begin{equation}
\frac{ge}{\hbar e}=\frac{n}{2}\ ,
\label{eq4}
\end{equation}
dove $n$ \`{e} un numero intero. Per $n=1$ otteniamo l'equazione fondamentale che lega le cariche fondamental $e$ e $g_D$. Se, invece, consideriamo un universo di dioni (in una teoria elettromagnetica), la condizione di quantizzazione (\ref{eq4}) deve essere sostituita dall'equazione\footnote{J. SCHWINGER; {\it Phys. Rev}, {\bf 144}, 1087 (1966); D. ZAWANZIGER; {\it Phys. Rev}., {\bf 176}, 1480, 1489 (1968).}
\begin{equation}
e_ig_j-g_ie_j=m(\hbar c),
\label{eq5}
\end{equation}
dove $(e_i,g_j)$ denotano le cariche elettriche e magnetiche del dione $i$ ed $m$ \`{e} un numero intero. Una piccola revisione di qualche punto basilare che ci porta all'eq. (\ref{eq4}) e l'interpretazione usuale dall'eq. (\ref{eq5}) sono presentate nela sez. 2. Nostro obiettivo in questa nota non \`{e} discutere il problema dell'esistenza teorica di un monopolo\footnote{E. AMALDI: \textit{On the Dirac magnetic poles}, in \textit{Old and New Problems in Elementary Particles}, a cura di G. Puppi (New York, N.Y., 1968). Vedi anche il recente articolo di rivista di J. PRESKILL; {\it Annu. Rev. Nucl. Part. Sci}., {\bf 34}, 461 (1984); G. GIACOMELLI; {\it Rivista Nuovo Cimento}, {\bf 7}, no. 12 (1984).}, ma unicamente mostrare che esiste un modo semplice di pervenire
all'eq. (\ref{eq5}) che, allo stesso tempo, pu\`{o} fare luce sull'origine di questa condizione di quantizzazione per le cariche in un
universo di dioni. L'argomento semiclassico che ci porta a una nuova interpretazione di questa condizione \`{e} presentato nella sez. 3 e le conclusioni nella sez. 4.

\section{La relazione di Dirac fra le cariche fondamentali; elettrica e magnetica.}

L'evidenza empirica della natura quantica della cariche elettriche, secondo Dirac$^{2}$, potrebbe essere compresa se ci fosse almeno un monopolo magnetico in natura. Le cariche elettrica e magnetica sarebbero legate dall'eq. (\ref{eq4}). Questo risultato \`{e} ottenuto considerando la meccanica quantistica di un elettrone in presenza di un monopolo, mantenendo la forma
dell'accoppiamento minimo per l'interazione tra i monopolo ed il campo elettromagnetico, usando la libert\`{a} di gauge del potenziale elettromagnetico ed
anche la monodromia della funzione d'onda di Schr\"{o}dinger$^{1}$. Per\`{o}, per fare questo, Dirac ha dovuto immaginare il monopolo come l'estremit\`{a}  di un lungo solenoide semiinfinito, il quale non pu\`{o} essere osservato empiricamente, in modo che questa rappresentazione del monopolo sia accettabile. Per quello che riguarda il nostro obiettivo in questa nota conviene presentare succintamente un interessante
argomento semiclassico\footnote{A.S. Goldhaber; \textit{Phys Rev. B}, {\bf 140}, 1407 (1965).}
  che mostra la possibilit\`{a} di giungere all'eq. (\ref{eq4}) a partire dallo studio della collisione tra una
particella di carica elettrica $E$ ed un monopolo magnetico stazionario di carica $G$. Infatti, se si considera la particella incidente lungo la direzione
$z$ con un parametro d'urto sufficientemente grande, si pu\`{o} computare la sua variazione di momento
angolare $\Delta L_z$ supponendo che, dopo l'interazione, la particella subisca una deflessione solo nella direzione solo nella direzione azimutale.
I calcoli sono molto semplici -- vedi $^{(1)}$ -- e il risultato trovato \`{e} che	$\Delta L_z = 2EG/c$, dove $c$ \`{e} la velocit\`{a} della luce. Pertanto, questa  quantit\`{a} \`{e} indipendente dal parametro  d'urto della collisione e dalla velocit\`{a} della particella e dipende solo dal prodotto $EG$.

Imponendo che la variazione $\Delta L_z$, sia un multiplo intero di $\hbar$, ne segue l'eq. (\ref{eq4}). \`{E} evidente che questo ragionamento presuppone  l'esistenza di un monopolo che contiene solo una carica magnetica. Tuttavia, per quello che ci proponiamo di discutere, bisogna supporre inizialmente che ogni particella possegga entrambe le cariche, sia quella elettrica, sia quella magnetica. Consideriamo due dioni con
cariche ($Q_ie$, $M_ig_D$) e ($Q_je$, $M_jg_D$)	misurate in unit\`{a} delle cariche elementari $e$ e $g_D$. L'idea di associare ad ogni dione un ``solenoide'' persiste e l'eq. (\ref{eq5}), che generalizza l'eq. (\ref{eq4}), pu\`{o} essere scritta come
\begin{equation}
Q_iM_j - Q_jM_i=m\ , \quad i\neq j\ ,
\label{eq6}
\end{equation}
dove $m$ \`{e} un intero. Questa equazione viene interpretata come la condizione necessaria e sufficiente che garantisce che un dione non possa rivelare il ``solenoide'' associato all'altro e viceversa. Vedremo adesso, nella prossima sezione, come si pu\`{o} giungere a questa condizione (\ref{eq6}) usando l'argomento semiclassico descritto sopra e l'invarianza delle equazioni di Maxwell per trasformazioni duali.

\section{Sulla condizione $Q_iM_j-Q_iM_i=m$.}

Supponendo un universo di dioni (nella teoria elettromagnetica di Maxwell) abbiamo visto nell'introduzione che questo universo sarebbe equivalente ad un altro dove non ci sono cariche e correnti magnetiche se, e soltanto se, il rapporto $G/E$ \`{e} universale. Dato che non sappiamo se ci\`{o} sia, partiamo dal presupposto che questo rapporto possa non essere universale. Cos\`{\i} scegliamo una trasformazione duale particolare, cio\`{e} $\mbox{tg}\,\theta =-Q_ie/M_ig$,
in modo che solo il dione $i$ si trasforma in un monopolo (cosa che \`{e} sempre possibile come si vede dalla (3)), cio\`{e}, le cariche iniziali ($Q_ie$, $M_ig$) diventano, con questa trasformazione ($0$, $M_ig/\cos \theta$). Chiamiamo questa particella $i$, ``monopolo di prova'',
il quale interagisce con un altro dione $j$, le cui
cariche ($Q_je$, $M_jg$)  si sono trasformate in ($Q^{\prime}_je$, $M^{\prime}_jg$), dove
\[
Q^{\prime}_j e =\cos \theta (Q_je+M_jg\ \mbox{tg}\, \theta)
\]
\[
M_j^{\prime}g =\cos \theta (-Q_je\ \mbox{tg}\, \theta +M_jg).
\]
Consideriamo il monopolo di prova in quiete ed il dione $j$ che si muove come descritto nella sez.~2. Poich\'{e} l'interazione tipo coulombiana tra le due cariche magnetiche \`{e} dovuta a una forza centrale, essa non alterer\`{a} il momento angolare della particella incidente. Pertanto permane valido il risultato precedente (sez. 2), dovuto all'interazione tra $E$ e $G$, dove $\Delta L_z = 2EG/c = m\hbar$. Nel nostro caso $E=Q^{\prime}_j$  e $G=M_ig/\cos \theta$. Cos\'{\i} ne segue che
\begin{equation}
\frac{eg}{\hbar c}=\frac{m}{2}\left(M_iQ_j-Q_iM_j\right)^{-1}
\label{eq7}
\end{equation}
Se misuriamo le cariche magnetiche in unit\`{a} di $g_D$, il primo membro della (\ref{eq7}) deve soddisfare l'eq. (\ref{eq4}) con $n=1$. Cos\'{\i}  l'uguaglianza
\begin{equation}
M_iQ_j-Q_iM_j=m
\label{eq8}
\end{equation}
\`{e} la condizione che garantisce che la relazione fondamentale di Dirac fra le due cariche elementari $e$ e $g_D$, sia mantenuta.

\section{Conclusioni.}

Abbiamo visto che \`{e} possibile usare l'argomento semiclassico presentato nella $^{(5)}$ in modo da giungere all'equazione $M_iQ_j-Q_iM_j=m$. Usando l'invarianza delle equazioni di Maxwell generalizzate per trasformazioni duali, si pu\`{o} sempre sceglierne una particolare, in modo da trasformare un universo composto soltanto di dioni in un altro che contiene almeno un monopolo. Dato che solo l'interazione di questo monopolo con le cariche elettriche degli altri dioni pu\`{o} portare ad  una variazione 
$\Delta L_z$	questo ``nuove universo'' di dioni pi\'{u} un monopolo \`{e} equivalente all'universo pensato da Dirac$^2$, dal punto di vista dell'argomento semiclassico che abbiamo usato.

Cio\`{e}, non importa se le particelle che interagiscono con il monopolo di prova sono elettroni $(M_i=0)$ o dioni. Pertanto, ci si deve aspettare che, se esistono due cariche fondamentali in natura ($e$ e $g_D$), la relazione fra loro deve essere preservata da una trasformazione duale. Cos\'{\i} il fatto che la condizione di quantizzazione per le cariche dei dioni non possa assumere valori semiinteri (come \`{e} il caso della condizione di Dirac) risulta dall'esigenza d'invarianza dell'equazione $eg_D=\hbar c/2$.

Abbiamo dunque visto che, oltre alla interpretazione usual secondo cui l'eq. (\ref{eq8}) \'{e} quella necessaria e sufficiente perch\'{e} un dione non possa rivelare il  ``solenoide'' di un altro, la condizione (\ref{eq8}) pu\`{o} essere interpretata come quella che infatti garantisce che la relazione fra le cariche elementari $e$ e $g_D$ sia universale.

\begin{center}
$\ast \ \ast \ \ast$
\end{center}

Vorrei ringraziare il prof. E. Predazzi  per la lettura critica di manoscritto e per i suggerimenti fatti, e anche il Conselho Nacional de Desenvolvimento Cientifico e Tecnol\'{o}gico CNPq, del Brasile, che ha finanziato questo lavoro.
\end{document}